\documentclass[twocolumn,showpacs,amsfonts,aps,prc,nofootinbib,floatfix,%
superscriptaddress]{revtex4}

\usepackage{amsmath}
\usepackage{bm}
\usepackage{graphicx}

\usepackage{epsfig}
\newcommand{\beq}{\begin{equation}}
\newcommand{\eeq}{\end{equation}}
\newcommand{\bea}{\vspace{0.25cm}\begin{eqnarray}}
\newcommand{\eea}{\end{eqnarray}}

\newcommand{\ta}{\mbox{{\boldmath
$\tau$}}}

\newcommand{\ro}{\mbox{{\boldmath
$\rho$}}}

\newcommand{\qb}{\mbox{{\bf
q}}}

\newcommand{\pt}{\mbox{{\bf
p}}_\perp}

\def\lsim{\mathrel{\rlap{\lower4pt\hbox{\hskip1pt$\sim$}}
    \raise1pt\hbox{$<$}}}         
\def\gsim{\mathrel{\rlap{\lower4pt\hbox{\hskip1pt$\sim$}}
    \raise1pt\hbox{$>$}}}         

\newcommand{\landau}{L.D.~Landau Institute for Theoretical Physics,
        GSP-1, 117940, Kosygina Str. 2, 117334 Moscow, Russia}

\begin{document}


\title{
Radiative quark $p_T$-broadening in a quark-gluon plasma beyond the 
soft gluon approximation
}
\date{\today}

\author{B.G.~Zakharov}\affiliation{\landau}

\begin{abstract}
We study the radiative correction to $p_T$-broadening of a fast quark 
in a quark-gluon plasma beyond the 
soft gluon approximation. 
We find that the radiative processes can suppress considerably 
$p_T$-broadening.
This differs dramatically from previous calculations to logarithmic 
accuracy in the soft gluon approximation, predicting
a considerable enhancement of $p_T$-broadening.

\end{abstract}
%

\maketitle
\noindent {\bf 1}.
Interaction of fast partons with quark-gluon plasma (QGP) leads to 
jet modification in $AA$-collisions.
It is dominated by radiative parton energy loss 
\cite{GW,BDMPS1,BDMPS2,LCPI1,GLV1,AMY,W1} due to parton multiple scattering in 
the QGP. The medium modification of jet fragmentation functions 
due to induced gluon emission leads to a strong suppression
of hadron spectra in $AA$ collisions at RHIC and LHC energies.
It is characterized by the nuclear modification factor $R_{AA}$.
In the last years the data on $R_{AA}$ from RHIC and LHC have been actively 
used for tomographic analyses of the QGP produced in $AA$ collisions.   
The suppression of particle spectra are related to modification 
of the jet parton distribution in the longitudinal (along the momentum of the
initial hard parton) fractional momentum. Multiple parton scattering 
in the QGP can also modify the transverse jet distribution due to
$p_{\perp}$-broadening of fast partons \cite{BDMPS2}. 
It should contribute to dijet and photon-jet angular decorrelation 
in $AA$ collisions.
Similarly to suppression of the hadron spectra, the observation of 
this effect could potentially give information on the density of the produced QCD matter.

For a single parton traversing a medium $p_{\perp}$-broadening
is usually characterized by the transport coefficient $\hat{q}$ 
\cite{BDMPS1,BDMPS2}: 
the mean squared momentum transfer for a gluon passing through 
a uniform medium of thickness $L$ is $\langle p_\perp^2\rangle=\hat{q}L$
(and for a quark $\langle p_\perp^2\rangle=\hat{q}LC_F/C_A$).
The radiative processes can give an additional 
contribution to $p_\perp$-broadening. The radiative contribution
to $\langle p_\perp^2\rangle$ has been addressed in recent
papers \cite{Wu_qhat-rad,Mueller_pt,Blaizot_pt}. It has been found
that the radiative contribution 
may be rather large. 
It mostly 
comes from the double logarithmic term $\ln^2(L/l_0)$ (where $l_0$ is about
the plasma Debye radius) \cite{Mueller_pt}. 
The analyses 
\cite{Wu_qhat-rad,Mueller_pt,Blaizot_pt}  have been performed in the
approximation of soft gluons. In the present letter we address
radiative $p_\perp$-broadening beyond the soft gluon approximation
and the logarithmic approximation used in \cite{Mueller_pt}.
We show that this reduces drastically the radiative contribution,
that can even become negative.
The analysis is based on the light-cone path integral (LCPI) 
\cite{LCPI1} approach. The general LCPI formulas for $p_\perp$ distribution
in a $a\to bc$ transition have been obtained in \cite{LCPI_PT} 
(see, also \cite{BSZ,Z_NP05}, and \cite{W1} in the soft gluon limit).

\begin{figure}
\includegraphics[height=2.5cm]{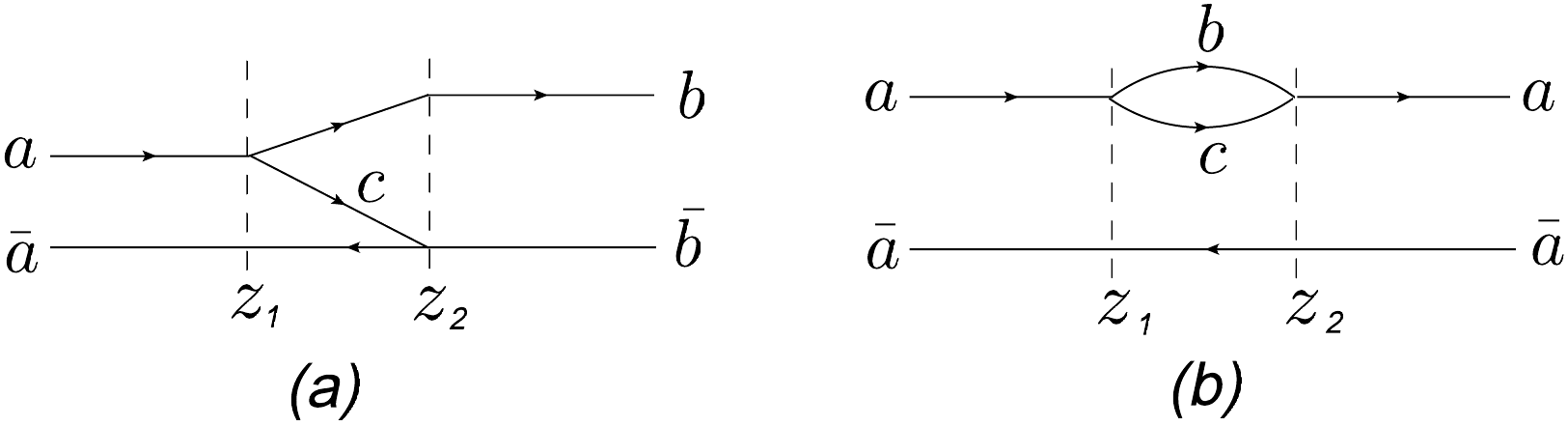}
\caption{\small Diagrammatic representation of ${dP}/{dx_bd\pt}$ ($a\to bc$
  process) (a) and of its virtual counterpart $d\tilde{P}/dx_bd\pt$
($a\to bc\to a$ process) (b). There are more two graphs
with interexchange of vertices between the upper and lower lines. 
}
\end{figure}

\noindent {\bf 2}.
We consider a fast quark with energy
$E$ produced at $z=0$ (we choose the $z$-axis along
the initial momentum of the quark) traversing a uniform medium of thickness
$L$. We account for only single gluon emission. Then, the final states 
include the quark and the quark-gluon system. We neglect collisional
energy loss (which is relatively small
\cite{Z_coll,Gale_coll}), then the energy of the
final quark without gluon emission equals $E$. 
In this approximation the medium does not change the energy 
for the one- and two-body states. The presence of the medium
modifies the relative fraction of the one-parton state and its transverse
momentum distribution, and for the two-parton channel the medium modifies 
both the longitudinal and transverse momentum distributions.
As in \cite{Mueller_pt,Blaizot_pt},  we will calculate 
the radiative correction to $p_\perp$-broadening of the final 
quark that includes
both the one- and two-parton channels, i.e., irrespectively to the 
longitudinal quark energy loss for the $qg$-state. 
In this formulation the radiative contribution
to $\langle p_\perp^2\rangle$  reads
\beq
\langle p_\perp^2\rangle_{rad}=\int dx_q d\pt \pt^2
\left[\frac{dP}{dx_qd\pt}+\frac{d\tilde{P}}{dx_qd\pt}\right]\,,
\label{eq:10}
\eeq
where $\frac{dP}{dx_qd\pt}$ is the distribution for 
real splitting $q\to qg$ in the transverse momentum of the final quark $\pt$
and its fractional longitudinal momentum $x_q$,
$\frac{d\tilde{P}}{dx_qd\pt}$ is the distribution
for the virtual process $q\to qg\to q$. In the latter case $x_q$ means 
the quark fractional momentum in the intermediate $qg$ system,
but $\pt$, as for the real process, corresponds to the final quark.  
The $x_q$-integration in (\ref{eq:10}) can equivalently be written
in terms of the gluon fractional momentum $x_g=1-x_q$. Below
we will denote $x_g$ as $x$.

Let us consider first the real splitting.
In the LCPI approach the distribution on the transverse momentum
and the longitudinal fractional momentum of the particle $b$ 
has the form \cite{LCPI_PT}
\beq
\frac{dP}{dx_b d\pt}=\frac{1}{(2\pi)^{2}}
\int\!\!
d\ta_f\,\exp(-i\pt\ta_f)F(\ta_f)\,,
\label{eq:20}
\eeq
where
\bea
F(\ta_f)=
2\mbox{Re}\!\!
\int_{0}^{\infty}\!\!dz_{1}\!\!\int_{z_{1}}^{\infty}\!\!dz_{2}
\Phi_{f}(\ta_f,z_{2})
\nonumber\\
\left.
\times\hat{g}
K(\ro_{2},z_{2}|\ro_1,z_{1})
\Phi_{i}(\ta_i,z_{1})\right|_{\ro_2=\ta_f, \ro_1=0}
\,,
\label{eq:30}
\eea
\beq
\Phi_{i}(\ta_i,z_{1})=
\exp\left[-\frac{\sigma_{a\bar{a}}(\ta_i)}{2}
\int_{0}^{z_{1}}\!dz n(z)\right]\,,
\label{eq:40}
\eeq
\beq
\Phi_{f}(\ta_f,z_{2})=
\exp\left[-\frac{\sigma_{b\bar{b}}(\ta_f)}{2}
\int_{z_{2}}^{\infty}\!dz n(z)\right]\,,
\label{eq:50}
\eeq
$\ta_i=x_b\ta_f$, 
$n(z)$ is the number density of the medium,
$\sigma_{a\bar{a}}$ and $\sigma_{b\bar{b}}$ are the dipole cross
sections for the $a\bar{a}$ and $b\bar{b}$ pairs,
$\hat{g}$ is the vertex operator,
$K$ is the Green function for the Hamiltonian
\beq
H=\frac{\qb^2+\epsilon^2}{2M}
-\frac{in(z)\sigma_{\bar{a}bc}(\ta_i,\ro)}{2}\,,
\label{eq:60}
\eeq
where $\qb=-i\partial/\partial \ro$, $M=E_{a}x_bx_c$, 
$\epsilon^2=m_{b}^{2}x_c+m_{c}^{2}x_b-m_{a}^{2}x_bx_c$
 with $x_c=1-x_b$,
and 
$\sigma_{\bar{a}bc}$ is the cross section for the 
three-body $\bar{a}bc$ system. The relative transverse parton positions 
for the $\bar{a}bc$ state read:
$\ro_{b\bar{a}}
=\ta_i+x_c\ro$,
$\ro_{c\bar{a}}
=\ta_i-x_b\ro$.
The vertex operator in (\ref{eq:30}) acts on the Green function as 
\bea
\hat{g}K(\ro_{2},z_{2}|\ro_1,z_{1})
=\frac{P^b_a(x_b)g(z_1)g(z_2)}{8\pi M^2}
\nonumber\\ \times
\frac{\partial}{\partial\ro_1}
\frac{\partial}{\partial\ro_2}
K(\ro_{2},z_{2}|\ro_1,z_{1})
\,,
\label{eq:70}
\eea
where $P^b_a(x_b)$ is the standard $a\to b$  splitting function
Note that the
derivatives on the right-hand side of (\ref{eq:70}) should be calculated 
for a fixed
$\ta_i$, i.e. for a fixed position of the center mass of the $bc$ pair.
%
The formula (\ref{eq:70}) is written for $z$-dependent 
coupling constant 
$g$, because the $z$-integrations in (\ref{eq:30}) extend up to
infinity, and the adiabatically vanishing coupling 
should be used. 

Diagrammatically, 
$\frac{dP}{dx_bd\pt}$ 
is shown in Fig. 1a. The
initial and final parallel   lines in Fig.~1 correspond to the Glauber 
factors (\ref{eq:40}) and (\ref{eq:50}), and the three-body
part between $z_1$ and $z_2$ corresponds to the Green 
function of the Hamiltonian (\ref{eq:60}). 
The factor $2$ in (\ref{eq:30}) accounts for the contribution from the diagram 
that can be obtained from Fig.~1a by interexchange of the vertices
between the upper and lower lines.
Diagram representation for the virtual process 
$a\to bc \to a$ that defines $\frac{d\tilde{P}}{dx_bd\pt}$ is shown in Fig.~1b.
In the virtual counterpart of (\ref{eq:30}) $\ta_i=\ta_f$, and the three-body
part 
also corresponds to the Green function (but now with 
arguments $\ro_1=\ro_2=0$) of the
Hamiltonian (\ref{eq:60}). The vertex factor for the virtual 
process changes sign.

For 
$q\to qg$ splitting (i.e. when $a=q$, $b=q$, $c=g$) the three-body cross section reads \cite{sigma3}
\beq
\sigma_{\bar{q}qg}=
\frac{9}{8}\left[\sigma_2(\ro_{qg})+\sigma_2(\ro_{g\bar{q}})\right]
-\frac{1}{8}\sigma_2(\ro_{q\bar{q}})\,,
\label{eq:80}
\eeq 
where $\sigma_2$ is the dipole cross section for the $q\bar{q}$ system.
We will use the quadratic approximation 
\beq
\sigma_2(\rho)=C\rho^2,\,\,\,\,\,\,C=\hat{q}C_F/2C_An\,.
\label{eq:90}
\eeq 
In this case the Hamiltonian (\ref{eq:60}) can be written in 
the oscillator form, 
for which one can use the analytical formula for the Green function.

At zero density 
the Glauber factors $\Phi_{i,f}$ become equal to unity, 
and the Green function is reduced to the vacuum one
\bea
K_{0}(\ro_2,z_2|\ro_1,z_1)=
\frac{M}{2\pi i(z_2-z_1)}
\nonumber\\ \times
\exp\left\{i\left[\frac{M(\ro_2-\ro_1)^{2}}{2(z_2-z_1)}-
\frac{\epsilon^2(z_2-z_1)}{2M}\right]\right\}\,.
\label{eq:100}
\eea
At any fixed $z_1$
\beq
\mbox{Re}
\int_{z_1}^{\infty}\!
dz_2
K_0(\ro_2,z_2|\ro_1,z_1)=
0\,.
\label{eq:110}
\eeq
However, the integration over $z_{1}$ in (\ref{eq:30}) is unconstrained,
and for a fixed coupling one gets the indeterminate product $0\times \infty$.
It can be resolved, using the exponentially decreasing coupling 
\beq
g(z)=g\exp(-\delta z)\,,
\label{eq:120}
\eeq 
and taking the limit $\delta \to 0$ after the $z_{1,2}$-integration.
This $\delta\to 0$ limit procedure gives for $n=0$ 
the standard spectrum for $q\to qg$ splitting
in vacuum
\beq
\frac{dP_0}{dx d\pt}=\frac{\alpha_sC_F}{2\pi^2x}\left[1+(1-x)^2\right]
\frac{\pt^2}{(\pt^2+\epsilon^2)^2}\,.
\label{eq:130}
\eeq

For a nonzero density the $z$-integrals in (\ref{eq:30}) over the region
$z_{1,2}>L$ also can be expressed via the vacuum spectrum. To separate
this contribution it is convenient to write
the product 
$\Phi_{f}(\ta_f,z_{2})
\hat{g}
K(\ro_{2},z_{2}|\ro_1,z_{1})
\Phi_{i}(\ta_i,z_{1})
$
in the integrand function on the right-hand side of (\ref{eq:30}) 
as
(we denote $\hat{g} K$ as ${\cal K}$ and omit arguments for 
notational simplicity)
\beq
\Phi_f{\cal K}\Phi_i=\Phi_f({\cal{K}}-{\cal{K}}_0)\Phi_i+
(\Phi_f-1){\cal K}_0\Phi_i
+{\cal{K}}_0(\Phi_i-1)+{\cal K}_0\,.
\label{eq:140}
\eeq
The last term on the right-hand side of (\ref{eq:140})
just corresponds to the vacuum splitting.
It can be omitted because it does not contain medium effects.

The Green function $K$ in the oscillator approximation, similarly to the
vacuum one (\ref{eq:100}), is the exponential of a quadratic form of 
the transverse
vectors $\ro_{1,2}$. 
In this case, each of the medium dependent terms in (\ref{eq:140}) 
is a combination of terms of the types
$\exp(-\ta_f^2A)$ and $\ta_f^2\exp(-\ta_f^2A)$, and, for given values of $z_{1,2}$ the $\ta_f$ integration in (\ref{eq:20}) becomes Gaussian. It allows one to 
represent the $\pt$-distribution (\ref{eq:20}) via the $z_{1,2}$-integrals
\cite{LCPI_PT}. 
However, for derivation of the $\langle p_\perp^2\rangle$
the explicit form of the $\pt$-distribution is unnecessary. Because
from (\ref{eq:20})
it is clear that it may be written as the Laplacian 
$\nabla^2$ of the function $F$ at $\ta_f=0$. The Laplacian for
the first three terms on the right-hand side of (\ref{eq:140}), that we need
for calculation of $\nabla^2F$ at $\ta_f=0$, 
read 
\bea
\nabla^2[\Phi_f({\cal K}-{\cal K}_0)\Phi_i]
=\nabla^2\Phi_f({\cal K}-{\cal K}_0)+({\cal K}-{\cal K}_0)\nabla^2\Phi_i
\nonumber \\ 
+\nabla^2({\cal K}-{\cal K}_0)\,,\,\,\,\,\,\,\,\,\,
\label{eq:150}
\eea
\beq
\nabla^2[(\Phi_f-1){\cal K}_0\Phi_i]
=\nabla^2\Phi_f{\cal K}_0\,,
\label{eq:160}
\eeq
\beq
\nabla^2[{\cal K}_0(\Phi_i-1)]
={\cal K}_0\nabla^2\Phi_i\,.
\label{eq:170}
\eeq

The total $\langle p_\perp^2\rangle_{rad}$ (\ref{eq:10}) includes also 
the contribution
of the virtual diagrams. As in (\ref{eq:10}) we will denote the quantities for
the virtual diagrams with a tilde. The real and virtual final Glauber
factors $\Phi_f$ and $\tilde{\Phi}_f$ are equal. Because they depend
on the $\ta_f$, which is same for real and virtual graphs.
For this reason the virtual contribution
will cancel the contributions for the real process 
in (\ref{eq:150}) and (\ref{eq:160}) that contain $\nabla^2\Phi_f$
(if we account for the fact that
 ${\cal K}=-\tilde{\cal K}$ and ${\cal K}_0=-\tilde{\cal K}_0$ 
at $\ta_f=0$). 
However, the terms with $\nabla^2\Phi_i$ in (\ref{eq:150}) and (\ref{eq:170}) 
are
not canceled by the contributions from the virtual diagrams.
Because the argument $\ta_i$ has different values 
for the initial state Glauber factors 
for the real and virtual splitting: $\ta_i=(1-x)\ta_f$ for the real
case and $\ta_i=\ta_f$ for the virtual one.
Then, the total
$\langle p_\perp^2\rangle_{rad}$,
corresponding to the sum $F+\tilde{F}$,
 can be written as 
\beq
\langle p_\perp^2\rangle_{rad}=I_1+I_2+I_3\,,
\label{eq:180}
\eeq
\bea
I_1=2\mbox{Re}\!\!\int\!dx\!\!\int_{0}^L\!\!\! dz_1\!\!
\int_{0}^{\infty}\!\!\!\!dz_{21}
\nabla^2({\cal K}\!-\!{\cal K}_0
+
\tilde{{\cal K}}\!-\!\tilde{{\cal K}}_0)\,,
\label{eq:190}
\eea
\bea
I_2=
2\mbox{Re}\!\int\!dx\!\!\int_{0}^L\!\!\! dz_1
\!\!\int_{0}^{\infty}\!\!\!dz_{21}\!
\left[
({\cal K}-{\cal K}_0)\nabla^2\Phi_i\right.
\nonumber\\
+
\left.(\tilde{{\cal K}}-\tilde{{\cal K}}_0)\nabla^2\tilde{\Phi}_i
\right]
\nonumber\\
=
-2\langle p_\perp^2\rangle_0
\mbox{Re}\!\int\!dx f(x)\!\!\int_{0}^L\!\!\! dz_1
\!\frac{z_1}{L}\int_{0}^{\infty}dz_{21}
({\cal K}-{\cal K}_0)\,,\,\,\,\,\,
\label{eq:200}
\eea
\bea
I_3=2\mbox{Re}\!\int dx\int_{0}^\infty dz_1\int_{0}^{\infty}dz_{21}
\left[{\cal K}_0\nabla^2\Phi_i+\tilde{{\cal K}}_0\nabla^2\tilde{\Phi}_i\right]
\nonumber\\
=-2\mbox{Re}\!\int\!dx f(x)\!
\int_{0}^\infty\!\!\!dz_1\int_{0}^{\infty}\!\!dz_{21}
{\cal K}_0\nabla^2\tilde{\Phi}_i\,\,\,\,\,\,\,
\label{eq:210}
\eea
with 
$
f(x)=x(2-x)\,$,
and $z_{21}=z_2-z_1$.
As in (\ref{eq:140})--(\ref{eq:170}), we omit arguments
for simplicity. 
In (\ref{eq:190})--(\ref{eq:210}) all the functions in the integrands 
should be calculated at $\ta_f=0$. 
The last lines in (\ref{eq:200}) and (\ref{eq:210}) used the fact 
that  at  $\ta_f=0$   
${\cal K}=-\tilde{\cal K}$, ${\cal K}_0=-\tilde{\cal K}_0$,
$\nabla^2\Phi_i=(1-x)^2\nabla^2\tilde{\Phi}_i$,
and  $\nabla^2\tilde{\Phi}_i$ equals $\langle p_\perp^2\rangle_0z_1/L$,
where 
$\langle p_\perp^2\rangle_0$ 
corresponds to nonradiative $p_\perp$-broadening.

The integrations over $z_1$ in (\ref{eq:190}) and (\ref{eq:200}) 
are constrained by $z_1=L$, because
${\cal K}-{\cal K}_0$  and
$\tilde{\cal K}-\tilde{\cal K}_0$ vanish at $z_{1}>L$. 
Note that it can be carried out setting $\delta=0$ in (\ref{eq:120}).
However, the integration over $z_{1,2}$ in (\ref{eq:210}), similarly to
calculation of the vacuum spectrum (\ref{eq:130}), is unconstrained,
and should be performed for a finite $\delta$, and then taking
the limit $\delta\to 0$. 
The $\delta\to 0$
limit procedure allows to represent (\ref{eq:210})
in the form
\beq
I_3=-\langle p_\perp^2\rangle_0 \!\int dx
f(x)\frac{dP_0}{dx}\,,
\label{eq:220}
\eeq
where 
\beq
\frac{dP_0}{dx}=\int d\pt
\frac{dP_0}{dx d\pt}
\label{eq:230}
\eeq
is the $\pt$-integrated
vacuum spectrum (\ref{eq:130}). The $\pt$-integral 
in (\ref{eq:230}) is logarithmically divergent.
This occurs because the formula (\ref{eq:20}) 
is obtained in the small angle approximation \cite{LCPI_PT}, 
and ignores the kinematic limits.  
We regulate (\ref{eq:230}) by restricting the integration region to 
$p_\perp<p_\perp^{max}$ with $p_\perp^{max}=E\mbox{min}(x,(1-x))$.
Formally, this divergence may be regulated by introduction of
the Pauli-Villars counter term with $\epsilon$ replaced by 
$\epsilon'\sim p_\perp^{max}$.

The $z_{21}$-integral in (\ref{eq:190}) is also logarithmically 
divergent, because
the integrand behaves as $1/z_{21}$ when $z_{21}\to 0$. 
Similarly to the logarithmic divergence of the $p_\perp$-integration for 
$I_3$, this
divergence is a consequence of the small angle approximation. 
And it also can be
regulated by the Pauli-Villars counter terms with $\epsilon'\sim p_\perp^{max}$.
Such counter terms will suppress the integrand at 
$z_{21}\lesssim M/\epsilon'^2$ (for small $x$ it is equivalent to 
$z_{21}\lesssim 1/\omega$). However, this 
would be reasonable only for a medium with a vanishing 
longitudinal correlation size. For the real QGP with the correlation
radius $\sim 1/m_D$ (here $m_D$ is Debye mass for the QGP)
the medium effect on the diagrams shown in Fig.~1 should vanish
when $z_{21}$ becomes small as compared to the Debye radius. 
For this reason it is reasonable to regulate the $z_{21}$-integral
in (\ref{eq:190}) by using the lower limit $z_{21}\sim 1/m_D$ (that is bigger
than $1/\omega$ at $\omega\gg m_D$).
This prescription has been used in \cite{Mueller_pt} for calculation
in the logarithmic approximation
of the contribution corresponding to our $I_1$  (\ref{eq:190}). It was 
found that the dominating contribution 
comes from the double logarithmic term $\propto \ln^2(L/l_0)$ with
$l_0$  the minimum $z_{21}$.
The contributions from 
$I_2$ and $I_3$ terms
have not been 
included in \cite{Mueller_pt}. As will be seen below,
these terms turn out to be very important, because they are negative 
and comparable to $I_1$. As a result, they change 
$\langle p_\perp^2\rangle_{rad}$ drastically.

\noindent {\bf 3}.
To make estimates of $\langle p_\perp^2\rangle_{rad}$
we use the quasiparticle masses $m_{q}=300$ and $m_{g}=400$ MeV  
\cite{LH}, that have been used in our previous 
analyses \cite{RAA13,RPP14}
of the RHIC and LHC data on the nuclear modification factor 
$R_{AA}$. 
The calculations of \cite{RAA13,RPP14} have been performed
for a more sophisticated model. In \cite{RAA13,RPP14} the induced gluon
emission has been calculated with running $\alpha_s$ 
for the QGP with Bjorken's longitudinal expansion,
which corresponds to $\hat{q}\propto 1/\tau$. 
In the present analysis, as in \cite{Mueller_pt}, we use constant 
$\hat{q}$ and $\alpha_s$.
To make our estimates as accurate as possible 
we adjusted the value of $\hat{q}$ to reproduce the quark
energy loss $\Delta E$ for running $\alpha_s$
in the model of \cite{RPP14} with the Debye mass from the lattice
calculations \cite{Bielefeld_Md}. 
As in \cite{Mueller_pt}, we take $\alpha_s=1/3$ and $L=5$ fm.
We obtained $\hat{q}\approx 0.27$ GeV$^3$ at $E=30$ GeV 
for Au+Au collisions
at $\sqrt{s}=0.2$ TeV and 
$\hat{q}\approx 0.32$ GeV$^3$ at $E=100$ GeV for Pb+Pb collisions
at $\sqrt{s}=2.76$ TeV.  

From the point of view of the numerical 
predictions for $\langle p_\perp^2\rangle_{rad}$ within 
the oscillator approximation (\ref{eq:90}), it is important
that the transport coefficient is an energy dependent quantity.
The energy dependence appears due to the Coulomb effects.
To a good approximation 
$\hat{q}$
can be written as \cite{BDMPS2,Baier_q,JET_q}
\beq 
\hat{q}=n\int_{0}^{p_{\perp max}^2} dp_\perp^2 p_\perp^2\frac{d\sigma}{d
  p_\perp^2}
\label{eq:240}
\eeq
with $p^2_{\perp max}\sim 3\omega T$, $\omega$ the gluon energy, 
$T$ the QGP temperature, $d\sigma/d p_\perp^2$ the differential gluon
cross section. 
The $p_\perp^2$-integration in (\ref{eq:240}) is logarithmic, and for this
reason $\hat{q}$ has a weak energy dependence.
The induced gluon emission is dominated by radiation
of soft gluons with $x\ll 1$. The typical gluon energy, $\bar{\omega}$,
is small compared to the initial quark energy, and depends 
weakly on $E$ \cite{Z_OA}. 
At $x\ll 1$  the induced 
gluon emission is dominated by the gluon multiple scattering.
For this reason, the induced gluon
spectrum is controlled by the value of the transport coefficient 
for soft gluons. Since the energy dependence of $\hat{q}$ is weak,
it can be calculated at $\omega\sim \bar{\omega}$. 
In the case of interest, 
$\bar{\omega}\sim 3-5$ GeV for a quark with $E\sim 30-100$ GeV. 
The above adjusted values of $\hat{q}$
correspond just to the transport 
coefficients for gluons with energy $\sim\bar{\omega}$.
However, the Glauber factors $\Phi_i$ and $\tilde{\Phi}_i$, that enter
(\ref{eq:200}) and (\ref{eq:210}), correspond to the initial quark,
and they should be calculated with the transport coefficient 
at energy $E$. We will denote it as $\hat{q}'$,
leaving the notation $\hat{q}$ for the transport coefficient at 
$\bar{\omega}$. 
Since $E\gg \bar{\omega}$, the ratio $r=\hat{q}'/\hat{q}$ 
may differ significantly from unity.
With the help of the formula (\ref{eq:240}) 
using the Debye mass from \cite{Bielefeld_Md} and
running $\alpha_s$ parametrized as in our previous 
jet quenching analysis 
\cite{RPP14} we obtained 
\beq
r\approx 2.4(2.63)
\label{eq:241}
\eeq
at $E=30(100)$ GeV for quark jets for RHIC(LHC) conditions.

In numerical calculations in (\ref{eq:200})--(\ref{eq:210}) 
we integrate over $x$ from $x_{min}=m_g/E$
to $x_{max}=1-m_q/E$.
As in \cite{Mueller_pt}, for the cutoff in the $z_{21}$-integration 
we use $z_{21}^{min}=1/m$ with $m=300$ MeV. 
For 
the three terms in (\ref{eq:180})
we obtained
\beq
[I_1,I_2,I_3]/\langle p_\perp^2\rangle_{0}\approx
[0.436/r,-0.213,-0.601]\,
\label{eq:250}
\eeq
at $E=30$ GeV for the RHIC conditions, and 
\beq
[I_1,I_2,I_3]/\langle p_\perp^2\rangle_{0}\approx
[0.85/r,-0.107,-0.908]
\label{eq:260}
\eeq
at $E=100$ GeV for the LHC conditions.
Using (\ref{eq:241}) we obtain 
from (\ref{eq:250}) and (\ref{eq:260})
for our RHIC(LHC) versions
\beq
\!\langle p_\perp^2\rangle_{rad}/
\langle p_\perp^2\rangle_{0}
\approx -0.632 (-0.692)\,,\,\,
r=2.4(2.63)\,.
\label{eq:270}
\eeq
And if we ignore the difference between $\hat{q}'$ 
and $\hat{q}$ 
\beq
\!\langle p_\perp^2\rangle_{rad}/
\langle p_\perp^2\rangle_{0}
\approx -0.378(-0.165)\,,\,\,
r=1(1)\,.
\label{eq:280}
\eeq
One sees that in all the cases
the radiative contribution to the mean squared $p_\perp$ is negative.  
This differs drastically from the prediction  of \cite{Mueller_pt}
$\langle p_\perp^2\rangle_{rad}\approx 0.75\hat{q}L$. In the form used 
in (\ref{eq:270}), (\ref{eq:280}) 
it reads
$\langle p_\perp^2\rangle_{rad}/
\langle p_\perp^2\rangle_{0}\approx 0.75\frac{C_A}{rC_F}\approx 1.7/r$.
The negative values of (\ref{eq:270}), (\ref{eq:280}) are due to 
a large negative contribution from $I_{2,3}$.
Since these terms 
have not been accounted for in \cite{Mueller_pt},  
it is interesting to compare prediction of \cite{Mueller_pt} 
with our results for $I_1$ term alone.
From (\ref{eq:250}) and (\ref{eq:260}) one can see that 
our $\left.\langle p_\perp^2\rangle_{rad}\right|_{I_1}$
is smaller than $\langle p_\perp^2\rangle_{rad}$
from \cite{Mueller_pt} by a factor of $\sim 3.9(2)$
for the RHIC(LHC) cases. This discrepancy says that the
logarithmic approximation used in \cite{Mueller_pt}
is rather crude.

Thus, we have found that the radiative contribution to $p_\perp$-broadening 
may be negative, or at least strongly suppressed as compared to
the predictions of \cite{Mueller_pt,Blaizot_pt}.
This seems to be supported by
the recent STAR measurement of the hadron-jet correlations \cite{STAR1},
in which no  evidence  for large-angle jet scattering in
the medium has been found.  
Similar to the analyses of \cite{Mueller_pt,Blaizot_pt}, our calculations
are performed for a uniform medium in the oscillator approximation. 
It would be interesting to perform calculations 
for an expanding QGP, and to go beyond the oscillator approximation.
We leave it for future work.
Of course, it is highly desirable to study the higher order effects.
However, even in the oscillator approximation and for a uniform medium, 
such calculations are  extremely difficult \cite{Arnold_2g1}.

This work has been supported by the RScF grant 16-12-10151.

\end{document}